\documentclass[10pt]{article}
\usepackage{graphicx, amssymb, amsmath}
\input epsf.tex

\let\badcite=\cite
\def\cite{~\badcite}

\def\slashchar#1{\setbox0=\hbox{$#1$}           
   \dimen0=\wd0                                 
   \setbox1=\hbox{/} \dimen1=\wd1               
   \ifdim\dimen0>\dimen1                        
      \rlap{\hbox to \dimen0{\hfil/\hfil}}      
      #1                                        
   \else                                        
      \rlap{\hbox to \dimen1{\hfil$#1$\hfil}}   
      /                                         
   \fi}
    \def\slashword#1{\setbox0=\hbox{$#1$}        
  \dimen0=\wd0                                   
   \setbox1=\hbox{/} \dimen1=\wd1                
   \ifdim\dimen0>\dimen1                         
      \rlap{\hbox to \dimen0{\hfil\bf---\hfil}} %
      #1                                         %
   \else                                         
      \rlap{\hbox to \dimen1{\hfil$#1$\hfil}}    
      /                                          
    \fi}                                         %

\catcode`@=11
\newdimen\vbigd@men                             

\def\vbig#1#2{{\vbigd@men=#2\divide\vbigd@men by 2%
   \hbox{$\left#1\vbox to \vbigd@men{}\right.\n@space$}}}
\catcode`@=12

\catcode`@=11
\def\citenum#1{\csname b@#1\endcsname}
\catcode`@=12

\begin{document}

\begin{flushright}
{SCUPHY-TH-08005}\\
\end{flushright}

\bigskip\bigskip

\begin{center}{\Large\bf\boldmath
Hidden Thresholds: Reconstructing NP Masses}
\end{center}
\bigskip
\centerline{\bf P. Huang, N. Kersting, H.H. Yang }
\centerline{{\it Physics Department, Sichuan University, P.R. China 610065}}
\bigskip

\begin{abstract}

The Hidden Thresholds technique reconstructs New Physics (NP) masses from two or more simultaneous NP decays at a hadron collider.
We show how this works in several MSSM examples:  $\widetilde{\chi}_{i}^0 \widetilde{\chi}_{j}^0  \to e^+ e^- \mu^+ \mu^- \widetilde{\chi}_{1}^0 \widetilde{\chi}_{1}^0$ ($i,j=2,3,4$) and
${\widetilde\chi_1}^\pm {\widetilde\chi_2}^0 \to {W^\pm}^*(\to \ell^\pm \nu) {\widetilde\chi_1}^0 ~ {Z^0}^*(\to \ell^\pm \ell^\mp) {\widetilde\chi_1}^0 $.
\end{abstract}


At a hadron collider such as the Tevatron at Fermilab or the Large Hadron Collider (LHC) at CERN, searches
for NP states beyond the Standard Model (SM) must take into account the fact that partonic center-of-mass (CM) energies at these machines
are not tunable, as they will be at a much anticipated $e^+ e^-$ linear collider, but vary continuously, in principle,
from zero to the maximum CM energy. Moreover, many NP models predict the production of a long-lived particle that is likely to escape the
 detectors, carrying away missing energy; production cross sections will thus not exhibit sharp resonances at the positions of NP masses, and we must consider other ways to measure these.

One well-studied avenue is to construct invariant mass distributions of final jet or leptonic momenta in exclusive channels and study their endpoints, these being analytical functions of NP masses\cite{Bachacou}. There are several caveats to this method however: the exclusive channel under study must somehow be identified or assumed, backgrounds must not interfere with endpoint measurement, and there may be some model-dependence in the method of fitting the endpoint on a 1-dimensional histogram. The  first caveat is most severe, especially in a model such as the Minimal Supersymmetric SM (MSSM), where the gluino and squarks decay via literally hundreds of possible decay chains.
Although SM backgrounds can be reduced by requiring a suitable number of hard jets and isolated leptons, NP backgrounds which may potentially shift the endpoint are much more challenging.

 Our claim is that two important features of NP particle production, when used together, can greatly boost the efficacy of the above endpoint method. First, if NP particles carry a new conserved charge, such as R-parity in the MSSM, they will be multiply-produced. We may, for example, consider inclusive decay chains of the form $\mathbb{X} \to A B \to \mathrm{jets} + \mathrm{leptons} $, where $A$ and $B$ are NP states arising from parent systems $\mathbb{X}$, the total invariant mass of these latter assuming any value from $m_\mathbb{X} = m_A + m_B$ (henceforth designated `threshold production')  all the way up to the maximum machine energy.
 Second, depending on $m_\mathbb{X}$ and the exact way in which $AB$ decay to the specified endstate,  invariant masses constructed from the final jet and lepton momenta attain extrema for certain kinematic configurations only.
  At threshold production, in particular, one special configuration will simultaneously maximize several invariant masses; collecting a large number of threshold decays, a 2-d or 3-d scatter plot of these invariants would exhibit a clustering around this `threshold point.' Yet threshold production is clearly only an infinitesimal possibility at a hadron collider and one might expect `above-threshold' events ($m_\mathbb{X} > m_A + m_B$) to hide the threshold point (hereafter called a `hidden threshold'). Contrary to this intuition, however, we find that, for some invariant mass combinations, the hidden threshold is geometrically fortified by above-threshold events, allowing
  us to directly measure invariant mass endpoints to constrain NP masses.
  In this Hidden Threshold (HT) technique,  model-dependence is thereby greatly reduced (the precise identity of $\mathbb{X}$ is irrelevant),
  backgrounds are less harmful(both by having the wrong correlations, and being spread out in more dimensions) and
  measurement of endpoints more straightforward.

\section{$\widetilde{\chi}_{i}^0 \widetilde{\chi}_{j}^0  \to e^- \tilde{e}^+(\to e^+ \widetilde{\chi}_{1}^0) ~  \mu^- \tilde{\mu}^+ (\to \mu^+ \widetilde{\chi}_{1}^0)$}

\begin{figure}[!htb]
\begin{center}
\includegraphics[width=6in]{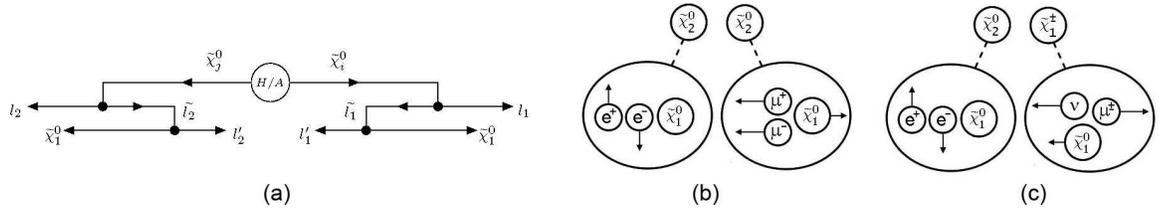}
\end{center}
\vskip -0.7cm
\caption{\small \emph{ Kinematic configurations of interest for (a) Neutralinos decaying through on-shell sleptons,
(b) off-shell sleptons, and (c) Neutralino-Chargino modes.}
}
 \label{fig:kin}
\end{figure}

\begin{figure}[!htb]
\begin{center}
\includegraphics[width=6in]{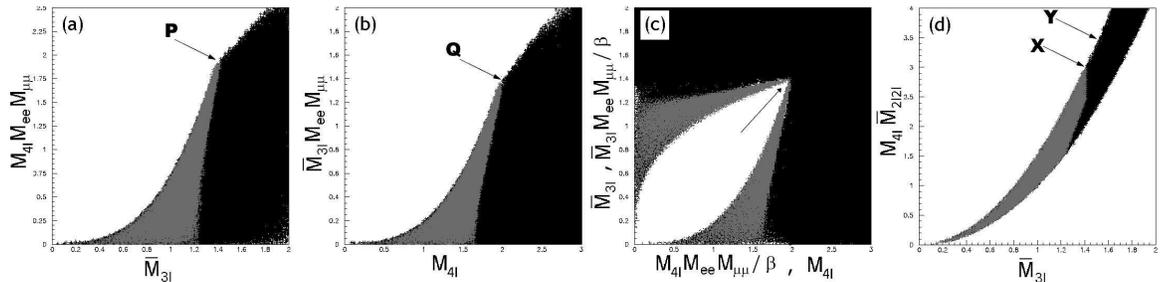}
\end{center}
\vskip -0.8cm
\caption{\small \emph{Invariant mass correlations for a superposition of Higgs decays with continuous mass
 $m_{\widetilde{\chi}_{i}^0} + m_{\widetilde{\chi}_{j}^0} < m_{H} < \infty$ (from relativistic kinematics only, with
$(m_{\widetilde{\chi}_{1}^0}, ~m_{\widetilde{\chi}_{i,j}^0}, ~m_{\tilde{e},\tilde{\mu}}) ~=~ (1,~2,~1.5)$ in arbitrary units). Points P and Q in (a) and (b) identify threshold endpoints; these can be found precisely with a `cat-eye' plot in (c), where we have abbreviated $\beta \equiv (M_{e e}^{max})\times(M_{\mu \mu}^{max})$. The position of
this threshold can now be marked (at X) in (d), with `pseudo-threshold' endpoints at any point Y  higher up on the envelope.
 }
}
 \label{fig:theory}
\end{figure}

 It will be easiest to explain HT by example, viz. neutralino-pair production in the  MSSM:
   $\mathbb{X} \to \mathbb{X}' + \widetilde{\chi}_{i}^0 \widetilde{\chi}_{j}^0 (\to e^+ e^- \mu^+ \mu^- 2
    \widetilde{\chi}_{1}^0)$ ($i,j = 2,3,4$) where $\mathbb{X}$  consists of some other
    pair of sparticles ($\tilde{g}\tilde{g}$, $\tilde{q}\tilde{g}$, $\tilde{q}\tilde{q}'$, $\tilde{q}\widetilde{\chi}_{i}^\pm$, etc.),
     a $Z^*$, or a heavy Higgs boson ($H^0$ or $A^0$),
     and $\mathbb{X}'$ is anything, e.g. hadronic jets, that does not confuse the 4-lepton signal.
      Let us explain the mechanism of HT in three steps:
in the first step we take $\mathbb{X}$ to be a single Higgs, the CM energy fixed
      at $m_{H}$;  the four leptons' momenta ($p_{1,2,3,4}$) can be systematically contracted into seven independent
      invariant masses\cite{Huang}, e.g.
        \begin{eqnarray}\label{avinv1}
  M_{4l}^2 &\equiv& (p_1 + p_{2}+ p_3 + p_{4})^2 \\ \nonumber
  \overline{M}_{{2l2l}}^4 &\equiv& \{(p_1 + p_{2}- p_3 - p_{4})^4 +
  (p_1 + p_{3}- p_2 - p_{4})^4 \\ \nonumber
  &&
  + (p_2 + p_{3}- p_{1} - p_{4})^4\} /3
     \\ \nonumber
  \overline{M}_{{3l}}^4 &\equiv&  \{
  (p_1 + p_{2}+ p_3)^4 +
  (p_1 + p_{2}+ p_{4} )^4 \\ \nonumber &&
  + (p_1 + p_{3}+ p_{4})^4 +
  (p_{2}+ p_{3}+ p_{4} )^4
 \}/4 \\ \nonumber
\end{eqnarray}
in addition to four others. The second step is to set the Higgs mass to a \emph{threshold} value, i.e. $m_{H} = m_{\widetilde{\chi}_{i}^0} + m_{\widetilde{\chi}_{j}^0}$. Now, it turns out, the invariants defined in (\ref{avinv1}), in addition to the usual dilepton invariants
$M_{ee}$ and $M_{\mu\mu}$, are \emph{simultaneously} maximal for the particular kinematic configuration shown in Fig.~\ref{fig:kin}a.
The final step is to take a continuous superposition of (hypothetical) Higgs' with masses ranging from the threshold value upwards, i.e. $m_{\widetilde{\chi}_{i}^0} + m_{\widetilde{\chi}_{j}^0} < m_{H} < \infty$.
 Fig.~\ref{fig:theory} shows what to expect for several choices of invariants: threshold decays are plotted in gray, while all above-threshold decays are plotted in black\footnote{Threshold decays obviously contribute only infinitesimally to the total shape, but for the
sake of seeing how these are distributed compared to above-threshold decays, we plotted $10^3$ of these on top of $10^6$ above-threshold events.}. The points P(Q) in Fig.~\ref{fig:theory}a(b) label
threshold endpoints and, intriguingly, are not obscured when above-threshold events are superimposed.
Obviously, without our color-coding in Fig.~\ref{fig:theory}, P and Q can only be said to lie on the envelope\footnote{There is a sort of kink near P and Q, but this turns out to be related to $M_{\ell^+ \ell^-}^{max}$ only.}, yet they must uniquely identify $M_{4l}^{max}$, $\overline{M}_{3l}^{max}$, and
$(M_{e e}^{max})\times(M_{\mu \mu}^{max})$, this last of which is obtainable from a wedgebox plot\cite{Bisset}, i.e. a plot of ``$M_{e e}~vs.~M_{\mu \mu}$".
We can therefore find the precise locations of P and Q with a graphical device:
superimpose a plot of ``$M_{4l}~vs.~\overline{M}_{3l}\frac{(M_{e e})\times(M_{\mu \mu})}{(M_{e e}^{max})\times(M_{\mu \mu}^{max})}$"  with one of ``$M_{4l}\frac{(M_{e e})\times(M_{\mu \mu})}{(M_{e e}^{max})\times(M_{\mu \mu}^{max})}~vs.~\overline{M}_{3l}$" (i.e. superimpose Fig.~\ref{fig:theory}a and Fig.~\ref{fig:theory}b with swapped and rescaled axes).
This gives rise to a characteristic 'cat-eye' shape(Fig.~\ref{fig:theory}c), where the upper corner of the eye pinpoints $M_{4l}^{max}$ and $\overline{M}_{3l}^{max}$.

There is another very useful correlation here, namely  ``$\overline{M}_{3l}~vs.~{M}_{4l} \overline{M}_{2l2l}$". Threshold values of $\overline{M}_{3l}^{max}$ and  $M_{4l}^{max} \overline{M}_{2l2l}^{max}$ again lie on the envelope at point `X' in Fig.~\ref{fig:theory}d, allowing us to solve for $\overline{M}_{2l2l}^{max}$; what is more interesting, every point on the envelope \emph{above} X (say, `Y' in Fig.~\ref{fig:theory}d) corresponds to another set of  $\overline{M}_{3l}^{max'}$ and  $M_{4l}^{max'} \overline{M}_{2l2l}^{max'}$ for a decay $\mathbb{X} \to \zeta \zeta$, where $\zeta$ is a hypothetical particle  which  decays just like a nuetralino ($m_{\zeta} > m_{\widetilde{\chi}_{i}^0}$).
This effectively says that the shape of the envelope above X depends on the masses $m_{\tilde{e},\tilde{\mu},\widetilde{\chi}_{1}^0}$ only, and precise measurement of this envelope can constrain these masses.

 Now there is really no difference between this situation with a continuously-massed Higgs and that of the general decay $\mathbb{X} \to \mathbb{X}' + \widetilde{\chi}_{i}^0 \widetilde{\chi}_{j}^0$. As long as the neutralino-pair subchains are intact, the mother chain is irrelevant. Note also we do not require the lightest neutralino to be stable, as long as its decay products do not include leptons.

\begin{figure}[!htb]
\begin{center}
\includegraphics[width=6in]{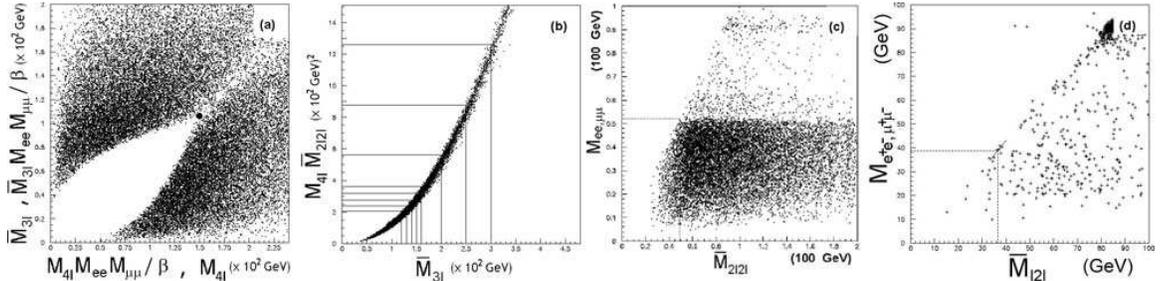}
\end{center}
\vskip -0.8cm
\caption{\small \emph{ Threshold values of $M_{4l}$ and $\overline{M}_{3l}$ for the On-Shell Point are found in the cat-eye plot (a) ($\beta = (75 \, \hbox{GeV})^2$) along with pseudo-threshold values on the envelope of the shape in (b). Simple graphical methods also give thresholds for off-shell topologies (c) and  chargino-neutralino modes (d). }
}
 \label{fig:mc}
\end{figure}

    Fig.~\ref{fig:mc}(a,b) show Herwig 6.5 simulations \footnote{For details on our MC setup and more, see \cite{Huang2}.} of Fig.~\ref{fig:theory}(c,d) for $30~fb^{-1}$ LHC luminosity at the following MSSM point: $(\tan\beta, ~\mu,~ M_1 | M_2,~M_{{(\tilde{e}, \tilde{\mu})}_L, \tilde{\tau}}, M_{{(\tilde{e}, \tilde{\mu})}_R},~M_A,~ M_{\tilde{q}} | M_{\tilde{g}})$ \newline $= (10,~250,~125 | 250,~250,~130,~700,~400 | 500)$ (masses in GeV). Here four-lepton endstates arise from colored sparticles  cascading down to a $\widetilde{\chi}_{2}^0 \widetilde{\chi}_{2}^0$-pair. Dilepton invariant masses thus have a sharp cutoff (here at $M_{e e, \mu \mu} \sim 75\, \hbox{GeV}$) which permits us to
    directly  construct the cat-eye plot in Fig.~\ref{fig:mc}a. Note that even at this low luminosity, the threshold values of  ${M}_{4l}^{max}$ and $\overline{M}_{3l}^{max}$(see black dot in Fig.~\ref{fig:mc}a) can be located to a few GeV precision.
    Several pseudo-threshold points for $\overline{M}_{3l} > \overline{M}_{3l}^{max}$ can now be measured in Fig.~\ref{fig:mc}b and fit to the parametric curve $(\overline{M}_{3l},{M}_{4l} \overline{M}_{2l2l})$ with
     { \small
\\
\\
\noindent\(
\mathbf{{M}_{4l}}= \frac{ m_\zeta^2(2 m_\zeta^2 m_s^2 - m_1^2  ) - m_s^4}{ m_j m_s^2}\)
\\
\\
\noindent\(
\mathbf{\overline{M}_{2l2l}}=
\frac{1}{3^{1/4}
  m_\zeta m_s^2} (3 m_1^8 m_\zeta^8 - 4 m_1^6 m_\zeta^6 m_s^2 (2 m_\zeta^2 + m_s^2) -
   4 m_1^2 m_\zeta^2 m_s^6 (8 m_\zeta^6 - 12 m_\zeta^4 m_s^2 + 6 m_\zeta^2 m_s^4 + m_s^6) +
   m_s^8 (16 m_\zeta^8 - 32 m_\zeta^6 m_s^2 + 24 m_\zeta^4 m_s^4 - 8 m_\zeta^2 m_s^6 +
      3 m_s^8) + 6 m_1^4 (4 m_\zeta^8 m_s^4 - 4 m_\zeta^6 m_s^6 + 3 m_\zeta^4 m_s^8))^{
 1/4}\)
\\
\noindent\(
\mathbf{\overline{M}_{3l}}= \frac{\left( m_1^4 m_\zeta^4 - 2 m_1^2 m_\zeta^4 m_s^2 +
    2 m_\zeta^4 m_s^4 - 2 m_\zeta^2 m_s^6 + m_s^8\right)^{1/4}
\sqrt{m_1^2 m_\zeta^2 - 2 m_\zeta^2 m_s^2 + m_s^4}}{2^{1/4} m_\zeta m_s^2}\)
\\
\\
}
($m_\zeta$ being the parameter), to constrain the LSP and slepton masses $(m_1, m_s)$.

\section{$\widetilde{\chi}_{i}^0 \widetilde{\chi}_{j}^0  \to e^- \tilde{e^*}^+(\to e^+ \widetilde{\chi}_{1}^0) ~  \mu^- \tilde{\mu^*}^+ (\to \mu^+ \widetilde{\chi}_{1}^0)$}

Now we take the sleptons to be off-shell, and the analysis simplifies. It turns out that the kinematic configuration in Fig.~\ref{fig:kin}b simultaneously maximizes $M_{e e}$ or $M_{\mu \mu}$ while minimizing $M_{4l}$, $\overline{M}_{2l2l}$,  and $\overline{M}_{3l}$:
\begin{equation}\label{offshell}
    \overline{M}_{4l,2l2l,3l}^{min} = (m_i-m_1)\left(
    \frac{\alpha + \beta \frac{m_1}{m_i}  + \gamma \frac{m_1^2}{m_i^2}  }{\xi}
         \right)^{\frac{1}{4}}
\end{equation}
where $(\alpha,\beta,\gamma,\xi) = (4,4,1,1), (2,0,1,3), (5,4,1,4)$, respectively.
 These, combined with the information in the dilepton endpoint, should yield mutually consistent
determinations of neutalino masses.
We now test this in the same MC setup as above at the following `Off-Shell Point' for $10~fb^{-1}$ LHC luminosity:
 $(\tan\beta, ~\mu,~ M_1 | M_2,~M_{{(\tilde{e}, \tilde{\mu})}_L, \tilde{\tau}},~M_{{(\tilde{e}, \tilde{\mu})}_R},~M_A,~ M_{\tilde{q}} | M_{\tilde{g}})$ \newline $= (10,~150,~100 | 200,~300,~135,~700,~300 | 350)$. The  relevant spectrum here is $(m_1, m_2, m_s) = (80, 132, 148)~ \hbox{GeV}$,
  forcing $\widetilde{\chi}_{2}^0$ to decay through an off-shell slepton to $\widetilde{\chi}_{1}^0$.
  Fig.~\ref{fig:mc}c allows us to directly read off, e.g., $\overline{M}_{2l2l}^{min}$ to good precision, and combining this with the dilepton endpoint constraint ($m_2 - m_1 = 52~\hbox{GeV}$) gives an accurate value of $m_1 = 85 \pm 11~\hbox{GeV}$.

\section{${\widetilde\chi_1}^\pm {\widetilde\chi_2}^0 \to {W^\pm}^*(\to \ell^\pm \nu) {\widetilde\chi_1}^0 ~~ {Z^0}^*(\to \ell^\pm \ell^\mp) {\widetilde\chi_1}^0 $}

As a final example, consider chargino-neutralino pairs which are forced to decay via off-shell gauge bosons (this is quite natural, for example, in Split-SUSY). Now the kinematic configuration in Fig.~\ref{fig:kin}c  maximizes $M_{\ell \ell}$ while minimizing another invariant:
\begin{equation}\label{ml2lmin}
    \overline{M}_{l2l}^{min} = \sqrt{m_{\widetilde{\chi}_{2}^0} - m_{\widetilde{\chi}_{1}^0}}
    \left( \frac{
    2(m_{\widetilde{\chi}_{2}^0} - m_{\widetilde{\chi}_{1}^0})^2
    + (m_{\widetilde{\chi}_{1}^\pm} - m_{\widetilde{\chi}_{2}^0} + m_{\widetilde{\chi}_{1}^0}
   -  m_{\widetilde{\chi}_{1}^0}^2 / m_{\widetilde{\chi}_{1}^\pm})^2  }{3}
       \right) ^{1/4}
\end{equation}
As in the last example, we have a simple graphical means of measuring this minimum, shown in Fig.~\ref{fig:mc}d for a typical Split-SUSY point \cite{Kersting}, which puts constraints on the masses of $\widetilde{\chi}_{1,2}^0$ and $\widetilde{\chi}_{1}^\pm$.

\section{Conclusions}

HT is a completely general procedure for any NP scenario: start with a decay chain of two or more NP particles to some observable jets and/or leptons, find kinematic configurations where some invariant masses are simultaneously extremal (and get analytical formulae for these extrema), and draw a 2d (or 3d) plot which makes these extrema measurable. Aside from the obvious applications to other sparticle-pairs, HT can be readily applied to other models with multiply-produced particles, e.g. extra dimensions, little Higgs, and lepto-quarks.


\vspace*{-.1in}
\begin{thebibliography}{99}
\vspace*{-.15in}

\bibitem{Bachacou}
H. Bachacou, I. Hinchliffe, and F.E. Paige,
``Measurements of masses in SUGRA models at CERN LHC'',
Phys.Rev.\textbf{D62}:015009 (2000).

\bibitem{Huang}
P. Huang, N. Kersting, and H.H. Yang,
``Extracting MSSM masses from heavy Higgs boson decays to four leptons at the CERN LHC'',
Phys.Rev.\textbf{D77}:075011 (2008).

\bibitem{Huang2}
P. Huang, N. Kersting, and H.H. Yang,
''Hidden Thresholds: A Technique for Reconstructing New Physics Masses at Hadron Colliders'',
arXiv:0802.0022 [hep-ph].

\bibitem{Bisset}
  M. Bisset et al.,
  `` Pair-produced heavy particle topologies: MSSM neutralino properties at the LHC from gluino/squark cascade decays'',
Eur.Phys.J.\textbf{C45}:477-492 (2006).

\bibitem{Kersting}
N. Kersting,
``On Measuring Split-SUSY Gaugino Masses at the LHC'',
arXiv:0806.4238 [hep-ph].


\end{thebibliography}
\end{document}